\documentclass{lcc}

\usepackage{amsmath}
\usepackage{amssymb}
\usepackage{epsfig}
\usepackage{enumerate}
\usepackage{ifthen}
\usepackage[margin=1.25in]{geometry}
\usepackage{url}

\newtheorem{theorem}{Theorem}
\newtheorem{corollary}[theorem]{Corollary}

\newtheorem{proposition}[theorem]{Proposition}

\newenvironment{proof}[1][]{{\emph{\ifthenelse{\equal{#1}{}}{Proof: }{Proof #1: }}}}{\hfill\scriptsize$\Box$}

\newcommand{\Omit}[1]{}
\newcommand{\denselist}{\topsep 0pt\itemsep 0pt}
\newcommand{\nat}{\mathbb{N}}
\renewcommand{\models}{\vDash}
\newcommand{\nmodels}{\nvDash}
\newcommand{\tup}[1]{\langle #1 \rangle}
\newcommand{\A}{\mathcal{A}}
\newcommand{\B}{\mathcal{B}}
\renewcommand{\L}{\mathcal{L}}
\newcommand{\struc}{\text{STRUC}}
\newcommand{\bit}{\text{BIT}}
\newcommand{\suc}{\text{suc}}
\newcommand{\FO}{\text{FO}}
\newcommand{\SO}{\text{SO}}
\newcommand{\MOD}{\text{MOD}}

\newcommand{\Inj}{\text{Inj}}
\newcommand{\F}{\mathcal{F}}
\newcommand{\C}{\mathbf{C}}
\newcommand{\Time}{\text{TIME}}

\newcommand{\fop}{\leq_{fop}}
\newcommand{\dual}[1]{\widehat{#1}}

\newcommand{\IS}{\textsc{IndependentSet}}
\newcommand{\CL}{\textsc{Clique}}
\newcommand{\SG}{\textsc{SubGraphIso}}

\begin{document}
\title{On Canonical Forms of Complete Problems via First-order Projections}
\author{Nerio Borges \\ Departamento de Matem\'aticas\\ Universidad Sim\'on Bol\'ivar\\ Caracas, Venezuela \\ \url{nborges@usb.ve} \and
        Blai Bonet \\ Departamento de Computaci\'on\\ Universidad Sim\'on Bol\'ivar\\ Caracas, Venezuela\\ \url{bonet@ldc.usb.ve}}
\maketitle

\begin{abstract}
The class of problems complete for NP via first-order reductions
is known to be characterized by existential second-order sentences
of a fixed form. All such sentences are built around the so-called
generalized IS-form of the sentence that defines \IS.
This result can also be understood as that every sentence that
defines a NP-complete problem $P$ can be decomposed in two disjuncts
such that the first one characterizes a fragment of $P$ as hard as \IS\ 
and the second the rest of $P$.
That is, a decomposition that divides every such sentence into a
a ``quotient and residue'' modulo \IS.

In this paper, we show that this result can be generalized over
a wide collection of complexity classes, including the so-called
nice classes. Moreover, we show that such decomposition can be
done for any complete problem with respect to the given class, 
and that two such decompositions are non-equivalent in general.
Interestingly, our results are based on simple and well-known
properties of first-order reductions.
\end{abstract}

\vskip 1eM
\noindent


\section{Introduction}

Descriptive complexity studies the interplay between complexity theory,
finite model theory and mathematical logic. 
Since its inception in 1974 \cite{fagin:spectra}, descriptive complexity
has been able to characterize all major complexity classes in term of logical
languages independent of any computational model, thus suggesting
that the computational complexity of languages is a property intrinsic
to them and not an accidental consequence of our choice for the computational
model.

In descriptive complexity, problems are understood as sets of (finite)
models which are described by logical formulae over given vocabularies.
Reductions between problems correspond to logical 
relations between the set of models that characterize the problems.
As important as the notion of polynomial many-one reductions in structural
complexity, there is the notion of first-order reductions in descriptive
complexity, and among such, the first-order projections (fops).
A fop is a very weak type of polynomial-time reduction whose study have
provided interesting results such as that common NP-complete problems like
\textsc{Sat}, \CL\  and others remain complete via fop reductions, and
that such NP-complete problems can be described by logical sentences in
a \emph{canonical form} \cite{dahlhaus:reduction,immerman:book}.

In this paper we continue the study of the syntactic aspects of complete
problems via fop reductions extending the work of Medina and Immerman
\cite{medina:syntactic,medina:thesis}.
In particular, we provide a general characterization of complete problems
via fops for a large collection of complexity classes that cover well beyond
just NP, including classes like P, PSPACE, $\Sigma^p_n$ and $\Pi^p_n$,
and others. Interestingly, our results rely on very general assumptions
and tools already known in the field.

The paper is organized as follows. In Sect.~2, we give
standard definitions and known results which provide
the theoretical framework of the paper and make it self contained.
Sect.~3 contains our main result, namely the generalization of
the Medina-Immerman result, together with relevant remarks and
some examples. Later, Sect.~4 shows a general result about the
existence of non-isomorphic problems via fop reductions, which
implies that our canonical form is indeed minimal.
Finally, Sect.~5 concludes with a brief summary and directions
for future work.

\section{Preliminaries}

\subsection{Logics, Finite Models, and Decision Problems}

A logical vocabulary is a tuple
$\tau=\tup{R_1^{a_1},\ldots,R_r^{a_r},c_1,\ldots,c_s,f_1^{r_1},\ldots,f_t^{r_t}}$
where the $R_j$s are relational symbols of arity $a_j$, $c_i$s are constant
symbols, and the $f_k$s are $r_k$-ary functional symbols. A structure for $\tau$,
also referred as $\tau$-structure or just structure if $\tau$ is clear from context,
is a tuple $\A=\tup{|\A|,R_1^\A,\ldots,R_r^\A,c_1^\A,\ldots,c_s^\A,f_1^\A,\ldots,f_t^\A}$
where
\begin{itemize}
\item[$\bullet$] $|\A|$ is the universe (or domain) of $\A$,
\item[$\bullet$] $R_j^\A \subseteq |\A|^{a_j}$ is a $a_j$-ary relation over $|\A|$,
\item[$\bullet$] $c_j\in|\A|$ is an element of the universe, and 
\item[$\bullet$] $f_k^\A:|\A|^{r_k}\rightarrow|\A|$ is a total $r_k$-ary function over $|\A|$.
\end{itemize}
For vocabulary $\tau$, $\struc[\tau]$ denotes the class of all finite structures,
i.e. those whose universe is an initial segment $\{0,1,\ldots,n-1\}$ of $\mathbb N$.

In addition to above logical symbols, we also have the \emph{numerical} relational
symbols `$=$', `$\leq$', `$\bit$' and `$\suc$', and constants `0' and `max', which
are assumed to belong to each vocabulary, and have \emph{fixed interpretations} on
every structure $\A$:
\begin{itemize}
\item[$\bullet$] $=$ and $\leq$ are interpreted as the usual equality and order on $\nat$,
\item[$\bullet$] $\A\models\bit(i,j)$ iff the $j$-th bit in the binary representation of $i$ is 1,
\item[$\bullet$] $\A\models\suc(x,y)$ iff $y$ is the successor of $x$ in the usual order on $\nat$, and
\item[$\bullet$] $0$ and $\max$ denote the least and greatest element in $|\A|$.
\end{itemize}
If $\L$ denotes a logic, the language $\L[\tau]$ is the set of all well-formed formulae
of $\L$ over the vocabulary $\tau$. A numerical formula in $\L[\tau]$ is a formula with
only numerical symbols. For example, $\SO\exists[\tau]$ is the set of all second-order
formulae of form $\exists Q_1\cdots\exists Q_n \Phi$ where the $Q_i$s are relational
variables and $\Phi$ is a first-order formula over over vocabulary $\tau$.
As usual, \FO\ denotes first-order logic and \SO\ denotes second-order logic.

A formula with no free variables is a sentence. For sentence $\varphi\in\L[\tau]$,
the class of all finite models that satisfy $\varphi$ is denoted as $\MOD[\varphi]$.
For fixed $\tau$, it is possible to code every finite $\tau$-structures into a sequence
of bits, i.e. a binary string, using a map $\MOD[\tau]\leadsto\{0,1\}^*$.
Hence, a collection of finite models can be represented as a collection of strings, 
or language.

In descriptive complexity, a decision problem $P$ is characterized by a subset of
models from $\struc[\tau]$ for some fixed $\tau$. For example, the problem \CL\
can be characterized by structures $\A=\tup{|\A|,E^\A,k^\A}$ over the vocabulary
$\tau=\tup{E^2,k}$, where $E$ is a binary relational symbol and $k$ is a constant,
such that $G=(|\A|,E^\A)$ makes up an undirected graph and $k^\A\in\{0,\ldots,|\A|-1\}$
denotes the size of a clique in $G$. Such models are typically characterized by a
sentence $\Psi$ over some fragment $\L$.  The problem \CL, for example, can be
characterized with a $\SO\exists$ sentence over $\tau$ \cite{fagin:spectra}; see below.

\subsection{First-Order Queries, Fops, and Duals}

Let $\tau$ and $\sigma$ be two vocabularies where
$\sigma=\tup{R_1^{a_1},\ldots,R_r^{a_r},c_1,\ldots,c_s}$ has no functional
symbols (from now on, we only consider vocabularies with no functional symbols).
Let $k\geq1$ and consider the tuple $I=\tup{\varphi_0,\ldots,\varphi_r,\psi_1,\ldots,\psi_s}$
of $r+s+1$ first-order formulae in $\FO[\tau]$ of form $\varphi_0(x_1,\ldots,x_k)$,
$\varphi_i(x_1,\ldots,x_{ka_i})$ and
$\psi_j(x_1,\ldots,x_k)=(x_1=c'_{j_1}\land\cdots\land x_k=c'_{j_k})$
where the $c'_{j_i}$s are constant symbols from $\tau$ (possibly with repetitions).
That is, $\varphi_0$ has at most $k$ free variables among $x_1,\ldots,x_k$,
$\varphi_i$ has at most $ka_i$ free variables among $x_1,\ldots,x_{ka_i}$,
and $\psi_j$ denotes a tuple in $\{c':c'\in\tau\}^k$.

Such tuple defines a mapping $\A\leadsto I(\A)$, called a \emph{first-order query}
of arity $k$, from $\tau$-structures into $\sigma$-structures given by:
\begin{itemize}
\item[$\bullet$] the universe $|I(\A)|\doteq\{(u_1,\ldots,u_k)\in|A|^k:\A\models\varphi_0(u_1,\ldots,u_k)\}$
                 is ordered lexicographically,
\item[$\bullet$] the relations are $R_i^{I(\A)}\doteq\{(\bar{u}_1,\ldots,\bar{u}_{a_i})\in|\A|^{ka_i}:\A\models\varphi_i(\bar{u}_1,\ldots,\bar{u}_{a_i})\}$,
\item[$\bullet$] the constants are $c_j^{I(\A)}\doteq\bar{u}$ for the unique $\bar{u}$ with $\A\models\varphi_0(\bar{u})\land\psi_j(\bar{u})$.
\end{itemize}
Furthermore, if for $T\subseteq\struc[\tau]$ and $S\subseteq\struc[\sigma]$, it is
true that $\A\in T$ iff $I(\A)\in S$, then $I$ is called a \emph{first-order reduction}
from $T$ to $S$.

\medskip

A first-order query is called a \emph{first-order projection} (fop) if
$\varphi_0$ is numerical and each $\varphi_i$ has form:
\[
\varphi_j(\bar x)\ \equiv\
  \alpha_0(\bar x)\ \lor\
    (\alpha_1(\bar x)\land\lambda_1(\bar x))\ \lor\ \cdots\ \lor\ (\alpha_r(\bar x)\land\lambda_r(\bar x))
\]
where the $\alpha_i$s are numerical and mutually exclusive, and each $\lambda_i$ is a $\tau$-literal.
Projections are typically denoted by the letter $p$.
If $p$ is a reduction from $\Pi$ to $\Psi$, we write $p:\Pi\fop\Psi$,
and if $\Pi$ is complete via $\fop$ reductions we say that $\Pi$ is $\fop$-complete.

Projections have interesting properties. For example, projections are a special
case of Valiant's non-uniform projections \cite{valiant:projections}.
For our purposes, we are interested in the fact that for each projection $p$ there
is a first-order sentence $\beta_p\in\FO[\sigma]$ that characterizes the image of $p$,
i.e. $\B\models\beta_p$ iff $\B=p(\A)$ for some $\A\in\struc[\tau]$.
The sentence $\beta_p$ is called the \emph{characteristic sentence}
of $p$ \cite{allender:isomorphism}.

\medskip

Finally, there is a syntactic operator associated to a first-order query that 
plays a fundamental role in our results.
For a first-order query $I:\struc[\tau]\rightarrow\struc[\sigma]$, the \emph{dual operator}
$\dual{I}$ maps formulae in $\L[\sigma]$ to formulae in $\L[\tau]$ in such a way that
\[ \A\models\dual{I}(\theta)\quad \text{if and only if}\quad  I(\A)\models\theta \]
for all $\theta\in\L[\sigma]$ and $\A\in\struc[\tau]$ \cite[Sect.~3.2]{immerman:book}.

\subsection{Complexity Classes}

For a family $\F$ of proper complexity functions \cite{papadimitriou:book}, we
consider the complexity classes $\Time(\F)=\cup_{f\in\F}\Time(f)$, and 
similarly for non-deterministic time and space. A complexity class $\C$ defined
by $\F$ is \emph{nice} if it has a universal problem of the form
\[ U_\C=\{\tup{M_i,\omega,1^t} : \text{$M_i$ accepts $\omega$ within $f_i(t)$ resources} \} \]
where $L(M_i)\in\C$ and $f_i\in\F$ bounds the resources of $M_i$.
Some well-known classes that are nice are L, NL, P, NP and PSPACE.
Allender et al. \cite{allender:isomorphism} showed that if $\Pi$ is
$\fop$-complete for a nice class $\C$, then it is complete via injective
fops of arity at least $2$. The following properties are shown easily:

\begin{proposition}
Let $\C$ be a complexity class defined by family $\F$. Then,
$(a)$ if $\C$ is a deterministic class and $\F$ is closed under sums,
then $\C$ is closed under finite unions;
$(b)$ if $\C$ is a nondeterministic class and $\F$ is such that for every $f,g\in\F$ there is
$h\in\F$ with $f,g\leq h$, then $\C$ is closed under finite unions;
$(c)$ if $\C$ is closed under finite unions and $\C$ is captured by logic $\L$, then
$\L$ is closed under disjunctions.
\end{proposition}

The nice classes L, NL, P, NP and PSPACE are known to be characterized
by SO-DetKrom, SO-Krom, SO-Horn, $\SO\exists$ and SO+TC respectively
\cite{gradel:lower,immerman:book}. Additionally, $\Sigma^p_k$ and
$\Pi^p_k$ are characterized by $\SO\exists\forall\cdots Q_k$ and
$\SO\forall\exists\cdots Q'_k$ sentences where $Q_k=\exists, Q'_k=\forall$
if $k$ is odd, and $Q_k=\forall, Q'_k=\exists$ if $k$ is even.
Thus, by the proposition, all these logical fragments are closed under
disjunctions, and also under conjunctions with first-order formulae.
We will make use of these facts later.

\section{Canonical Forms of Complete Problems}

Medina and Immerman characterized $\fop$-complete problems for NP syntactically
using the \IS\ problem. This problem consists of checking whether an input graph
$G$ has an independent set of size $k$. \IS\ is known to be complete for NP under
different notions of reductions, and in particular, under fop reductions \cite{medina:syntactic}.
\IS\ in characterized by the following $\SO\exists[\tau]$ sentence,
for $\tau=\tup{E^2,k}$:
\begin{equation}
\label{eq:is}
\Psi_{IS}\ =\ (\exists f\in\Inj)(\forall x,y)
   \bigl[x\neq y\,\land\,f_x\leq k\,\land\,f_y\leq k\,\rightarrow\,\neg E(x,y)\bigr]
\end{equation}
where `$f\in\Inj$' means that $f$ is a total and 1-1 function, i.e. an ordering of
the elements of the universe, and $f_x$ denotes $f(x)$. Although it seems that 
\eqref{eq:is} quantifies over a functional variable, $f$ is indeed a relational
variable such that $f_x$ is the unique element such that $f(x,f_x)$.
The condition $f\in\Inj$ is easily defined in first-order logic.
Observe that the only second-order variable in \eqref{eq:is} is $f$ which is
existentially quantified.

\begin{theorem}[\cite{medina:syntactic}]
\label{thm:medina}
Let $L\subseteq\struc[\sigma]$ be a NP problem characterized by $\Psi\in\L[\sigma]$
where $\sigma=\tup{Q^1}$ is the vocabulary of binary strings.
Then, a problem $L$ is NP-complete via $\fop$ reductions iff there is an injective
fop $p:\struc[\tup{E^2,k}]\rightarrow\struc[\sigma]$ such that
\begin{equation}
\label{eq:imm}
\Psi\ \equiv\ (\beta_p\,\land\,\Upsilon_{IS})\,\lor\,(\neg\beta_p\,\land\,\Lambda)
\end{equation}
where $\beta_p\in\FO[\sigma]$ is the characteristic sentence of $p$,
$\Upsilon_{IS}\in\SO\exists[\sigma]$ is a \emph{generalized IS-form}
\cite{medina:syntactic}, and $\Lambda$ is a $\SO\exists[\sigma]$ sentence.
\end{theorem}

Intuitively, this result says that if sentence $\Psi$ characterizes a
$\fop$-complete problem $L$ for NP, then it can be decomposed in two disjuncts
$\Psi=\Psi_{IS}\lor\Psi_{rest}$ such that $\MOD[\Psi_{IS}]$ is $\fop$-complete
for NP and $\MOD[\Psi_{rest}]$ equals the ``rest'' of $L$ which is not necessarily
complete.

Our main contribution is to show that above result can be generalized over
a wide collection of complexity classes, including the nice classes,
and that such decomposition can be done modulo any $\fop$-complete problem for
the given class.
Moreover, we also show two such decompositions are not in general equivalent.

The main obstacle for such generalization is to take care of the sentence
$\Upsilon_{IS}$ for classes different than NP. As it will be shown, we do not have
to consider each different class in isolation, since the corresponding
$\Upsilon$ sentences will be the duals of the sentence $\Psi$ that characterize
the complete problem.

Let us first define the relation $\cong_\Pi$ over $\struc[\tau]$ with respect to
a given problem $\Pi\subseteq\struc[\tau]$. For structures $\A$ and $\B$, define
\begin{equation}
\label{eq:eqrel}
\A\cong_\Pi\B\quad \text{iff}\quad (\A\in\Pi \Leftrightarrow \B\in\Pi)\,.
\end{equation}
Clearly, $\cong_\Pi$ is an equivalence relation that partitions
$\struc[\tau]$ into $\Pi$ and its complement.

By using dual operators and the equivalence relation, we are able to
show the following generalization of Theorem~\ref{thm:medina}.
In the following, $\tau$ and $\sigma$ refer to any two vocabularies.

\begin{theorem}[Main]
\label{thm:main}
Let $\C$ be a complexity class captured by fragment $\L$ closed under disjunctions
and closed under conjunctions with \FO. Let $\Pi\subseteq\struc[\tau]$ be a
$\fop$-complete problem for $\C$ characterized by $\Psi\in\L[\tau]$, and $B$ a
problem over vocabulary $\sigma$.
Then, $B$ is $\fop$-complete for $\C$ if and only if there is a fop
$p:\struc[\tau]\rightarrow\struc[\sigma]$ such that for all $\B\in\struc[\sigma]$:
\begin{equation}
\label{eq:thm}
\B \in B \quad\text{iff}\quad \B \models 
            (\beta_p\,\land\,\dual{I}(\Psi))\,\lor\,(\neg\beta_p\,\land\,\Lambda)
\end{equation}
where
\begin{enumerate}[{\rm($a$)}]\denselist
\item $\beta_p\in\FO[\sigma]$ is the characteristic of $p$,
       i.e. $\B\models\beta_p$ iff $\B\in p(\struc[\tau])$,
\item $\Lambda\in\L[\sigma]$, and
\item $I:\struc[\sigma]\rightarrow\struc[\tau]$ is a first-order query 
      such that for all $\A\in\struc[\tau]$, $I(p(\A))\cong_\Pi\A$.
\end{enumerate}
\end{theorem}
\begin{proof}
For the necessity, assume that $B$ is $\fop$-complete for $\C$; i.e.  $B$ is
characterized by some sentence $\Lambda\in\L[\sigma]$ and there is $p:\Pi\fop B$.
For $\B\in B$ we consider the two cases whether $\B\not\in p(\struc[\tau])$ or not.
For the first case, $\B\models\neg\beta_p\land\Lambda$.
For the second case, $\B\models\beta_p$ and
\begin{alignat*}{2}
 &\B=p(\A)                      &\quad\quad&\text{(for some $\A\in\struc[\tau]$ by $(a)$)} \\
 &\implies\ \A\in\Pi                      &&\text{(since $p$ is reduction)} \\
 &\implies\ \A\models\Psi                 &&\text{($\Psi$ characterizes $\Pi$)} \\
 &\implies\ I(p(\A))\models \Psi          &&\text{(by condition $(c)$)} \\
 &\implies\ p(\A)\models\dual{I}(\Psi)    &&\text{(def. of dual of $I$)}\,.
\end{alignat*}
Therefore, $\B\in 
B\implies\B\models(\beta_p\land\dual{I}(\Psi))\lor(\neg\beta_p\land\Lambda)$.
Now, let $\B\in\struc[\sigma]$ be such that
$\B\models(\beta_p\land\dual{I}(\Psi))\lor(\neg\beta_p\land\Lambda)$.
If $\B\models\Lambda$, then $\B\in B$. Otherwise,
\begin{alignat*}{2}
 &\B\models\beta_p\land\dual{I}(\Psi) \\
 &\implies\ \B=p(\A)\;\text{and}\;p(\A)\models\dual{I}(\Psi) &\quad\quad&\text{(for some $\A\in\struc[\tau]$)} \\
 &\implies\ I(p(\A))\models\Psi                              &&\text{(def. of dual)} \\
 &\implies\ \A\models\Psi                                    &&\text{(by $(c)$)} \\
 &\implies\ \A\in\Pi                                         &&\text{($\Psi$ characterizes $\Pi$)} \\
 &\implies\ \B \in B                                         &&\text{(since $p$ is reduction)}\,.
\end{alignat*}
It remains to show that there are first-order queries satisfying ($c$).
Since $\Pi$ is complete, there is a fop $I:\struc[\sigma]\rightarrow\struc[\tau]$
that reduces $p(\Pi)$ to $\Pi$. Note that $p(\Pi)\subseteq B$ since $p$ is
also a reduction. For $\A\in\struc[\tau]$, observe
\begin{eqnarray*}
&\A\in\Pi\ \Rightarrow\ p(\A)\in p(\Pi)\ \Rightarrow\ I(p(\A))\in\Pi\,, \\ 
&I(p(\A))\in\Pi\ \Rightarrow\  p(\A)\in p(\Pi)\ \Rightarrow\ p(\A)\in B \Rightarrow \A\in\Pi\,.
\end{eqnarray*}
Thus, $I:p(\Pi)\fop\Pi$ satisfies $\A\in\Pi$ iff $I(p(\A))\in\Pi$; i.e.  $\A\cong_\Pi I(p(\A))$.

For the sufficiency, assume there is a fop $p:\struc[\tau]\rightarrow\struc[\sigma]$
such that \eqref{eq:thm} holds for all $\B\in\struc[\sigma]$.
We need to show that $B$ is complete for $\C$. The inclusion $B\in\C$ is direct
from the closure properties on $\L$. For the hardness, we show that $p$ is indeed
a reduction from $\Pi$ to $B$.
For $\A\in\struc[\tau]$, we have $p(\A)\models\beta_p$. If $\A\in\Pi$, then
\[ \A\models\Psi\ \Rightarrow\ I(p(\A))\models\Psi\
                  \Rightarrow\ p(\A)\models\dual{I}(\Psi)\
                  \Rightarrow\ p(\A)\in B\,. \]
On the other hand, if $p(\A)\in B$, then
\[ p(\A)\models\beta_p\ \Rightarrow\ p(\A)\models\dual{I}(\Psi)\
                        \Rightarrow\ I(p(\A))\models\Psi\
                        \Rightarrow\ \A\models\Psi\ \Rightarrow\ \A\in \Pi\,.\]
Thus, $\A\in\Pi$ iff $p(\A)\in B$, $p$ is a reduction, and $B$ is complete.
\end{proof}

\begin{corollary}
The theorem holds if the first-order query $I$ is the reduction
$I:p(\Pi)\fop\Pi$ which exists since $\Pi$ is complete.
\end{corollary}

Moreover, a first-order query $J$ satisfying ($c$) is essentially equivalent
(with respect to $\Psi$) to the reduction $I:p(\Pi)\fop\Pi$.  Indeed, for such
$J$ and a finite $\sigma$-structure $\B=p(\A)$ for $\A\in\struc[\tau]$,
\[\B\models\dual{J}(\Psi) \iff J(\B)\models\Psi \iff \A\models\Psi \iff
  I(\B)\models\Psi \iff \B\models\dual{I}(\Psi)\,.\]

If we consider nice complexity classes, then the fop $p$ can be assumed to be
injective by a result of Allender et. al \cite{allender:isomorphism}.

\begin{corollary}
\label{cor:nice}
For nice classes, the fop $p:\struc[\tau]\rightarrow\struc[\sigma]$
can be assumed to be injective.
\end{corollary}

To see that Theorem~\ref{thm:medina} is equivalent to Corollary~\ref{cor:nice}
when $\C=\text{NP}$, let $\tau=\tup{E^2,k}$ and $\sigma=\tup{Q^1}$ be the
vocabularies for graphs and binary strings respectively, and consider
a problem $L\subseteq\struc[\sigma]$ complete for NP characterized by $\Psi_L$.
According to Theorem~\ref{thm:medina},
\[\Psi_L\ \equiv\ (\beta_p \land \Upsilon_{IS}) \lor (\neg\beta_p \land \Lambda)\]
where $p:\IS\rightarrow L$ is a first-order projection and $\Lambda$ is a
$\SO\exists$ sentence. On the other hand, according to Corollary~\ref{cor:nice},
$\Psi_L$ also satisfies
\[\Psi_L\ \equiv\ (\beta_p \land \dual{I}(\Psi_{IS})) \lor (\neg\beta_p \land \Lambda')\,.\]
As shown before, $\Upsilon_{IS}$ and $\dual{I}(\Psi_{IS})$ are equivalent on
$p(\struc[\tau])$, and thus $\Lambda$ and $\Lambda'$ must be equivalent on
$\struc[\sigma]\cap\text{MOD}[\neg\beta]$.

\subsection{Examples}

Consider $\CL\subseteq\struc[\tau=\tup{E^2,k}]$ characterized by the $\SO\exists$ sentence
\begin{displaymath}
\Psi_{CL}\ =\ (\exists f\in\Inj)(\forall x,y)
   \bigl[x\neq y\,\land\,f_x\leq k\,\land\,f_y\leq 
k\,\rightarrow\,E(x,y)\bigr]\,.
\end{displaymath}
For $\sigma=\tau$, it is not hard to see that \IS\ can be reduced to \CL\ using the fop
$p=\lambda_{xy}\tup{\varphi_0,\varphi_1,\psi}$, of arity 1, where
\[\varphi_0(x)\ =\ true\,, \quad \varphi_1(x,y)\ =\ \neg E(x,y)\,, \quad \psi(x)\ =\ (x=k)\,.\]
Clearly, if $\A=\tup{|\A|,E^\A,k^\A}$, then $|p(\A)|=|\A|$, $E^{p(\A)}=|\A|^2\setminus E^\A$
and $k^{p(\A)}=k^\A$. Therefore, $p(p(\A))=\A$ for all $\A\in\struc[\tau]$, and hence
\[p(p(\A))\in\IS\quad \text{iff}\quad \A\in\IS\,.\]
Furthermore, $\beta_p=true$ and since \CL\ is also known to be NP-complete with respect
to $\fop$ reductions, we have
\[\Psi_{CL}\ \equiv\ 
   (\beta_p\,\land\,\dual{p}(\Psi_{IS}))\,\lor\,(\neg\beta_p\,\land\,\Gamma)\ =\ \dual{p}(\Psi_{IS})\,.\]
Conversely, beginning with the observation tha $\beta_p=true$ and $\dual{p}(\Psi_{IS})=\Psi_{CL}$
we can conclude, by Theorem~\ref{thm:main}, that \CL\ is $\fop$-complete for NP.
We call this formulation of \CL\ as its canonical form with respect to \IS.
In this example, the formula $\Psi_{CL}$ was already in its canonical form
with respect to \IS.

\bigskip
For a second example, consider the problem \SG\ defined by tuples $\tup{G,G'}$ such that
the graph $G$ contains a subgraph isomorphic to graph $G'$. Such tuples can be expressed
with the vocabulary $\sigma=\tup{F^2,H^2,k}$ where $F$ and $H$ define the edges of $G$ and
$G'$, and the constant $k$ defines the initial segment $\{0,\ldots,k-1\}$ for the edges of $G'$.
Among other things, instances of \SG\ are identified with structures $\B$ in which
$H^\B\subseteq\{0,\ldots,k-1\}^2$. \SG\ is defined by the $\SO\exists$ sentence $\Psi_{SG}$
\begin{displaymath}
(\exists f\in\Inj)(\forall x,y)
   \bigl[x\neq y\,\land\,f_x<k\,\land\,f_y<k\,\rightarrow\,(H(f_x,f_y)\rightarrow F(x,y))\bigr]\,.
\end{displaymath}
A fop reduction $p$ from \CL\ into \SG\ outputs $\tup{G,K_k,k}$ on input $\tup{G,k}$.
The fop is $p=\tup{\varphi_0,\varphi_1,\varphi_2,\psi}$ given by
\[ \varphi_0\,=\,true\,,\ \varphi_1\,=\,E(x,y)\,,\ \varphi_2\,=\,(x<k \land y<k)\,,\ \psi\,=\,(x=k)\,.\]
The characteristic sentence of $p$ is
\[\beta_p\ =\ x<k\land y<k\rightarrow F(x,y)\,.\]
The reduction $I:p(\CL)\fop\CL$ given by $I=\tup{\varphi_0=true,\varphi_1=F(x,y)}$
satisfies $\B\in p(\CL)$ if and only if $I(\B)\in\CL$ for all $\B$. Since $\Psi_{SG}$ is equivalent
to $(\beta_p\land \dual{I}(\Psi_{CL}))\lor(\neg\beta_p\land \Psi_{SG})$, then,
by Corollary~\ref{cor:nice}, \SG\ is complete for NP via $\fop$ reductions.

\bigskip
Finally, other classes that satisfies the conditions of Corollary~\ref{cor:nice}
are L, NL, P, PSPACE, and all $\Sigma^p_k$ and $\Pi^p_k$.

\section{Non-Isomorphic Complete Problems for Nice Classes}

The next result is a more general version of one already known for NP \cite{medina:syntactic}.
The proof is analogous to the NP case. Among other things. it implies that we cannot
get rid of the disjunction in Corollary~\ref{cor:nice}.
\begin{theorem}\label{propnoiso}
If $\C$ is a nice complexity class, then there are two $\C$-complete
problems that are not fop-isomorphic.
\end{theorem}
\begin{proof}
Let $\Gamma\subseteq\{0,1\}^*$ be a $\fop$-complete problem for $\C$, and define
$\Gamma'=\{\omega0,\omega1:\omega\in\Gamma\}$.
It is easy to see that $\Gamma'$ is complete via fops; e.g. define the projection
$p:\struc[\tau=\tup{S^1}]\rightarrow\struc[\sigma=\tup{T^1}]$, of arity 2, as
$p=\tup{\varphi_0(x,y),\varphi_1(x,y)}$ where $\varphi_0(x,y)=(x=0)\lor(x=1\land y=0)$
gives the domain of $p(\A)$ and $\varphi_1(x,y)=(x=0\land S(y))\lor(x=1\land y=0)$
gives $T^{p(\A)}$.
Thus, for $\A$ with domain $|\A|=\{0,\ldots,n-1\}$, $\varphi_0$ defines
\[|p(\A)|\ =\ \{(0,y):0\leq y<n\}\,\cup\,\{(1,0)\}\,.\]
Formula $\varphi_1$ identifies the $n$ bits of $\A$ with the tuples $(0,x)$
and assigns ``value'' 1 to the tuple $(1,0)$. Observe that the order induced
in $p(\A)$ is $(0,0)<(0,1)<\cdots<(0,n-1)<(1,0)$.
Therefore, $\omega\in\Gamma$ iff $p(\omega)\in\Gamma'$ which shows that
$\Gamma'$ is complete.

Since $\C$ is a nice complexity class, there is a fop $p:\struc[\tau]\rightarrow\struc[\sigma]$
that is injective, of arity $k\geq 2$, that reduces $\Gamma$ to $\Gamma'$.
We will show that $p$ cannot be onto by showing that if $\omega\in\Gamma$, then
either $\omega0\not\in p(\Gamma)$ or $\omega1\not\in p(\Gamma)$.

Consider the formula $\varphi(\bar x)$ that defines the interpretation of $T$
in the structure $p(\A)$ of form
\[\varphi(\bar x)\ =\ \alpha_0(\bar x)\,\lor\,(\alpha_1(\bar x)\land\lambda_1(\bar x))\,\lor\cdots\lor\,(\alpha_r(\bar x)\land\lambda_r(\bar x))\,.\]
We are going to show $w0\in p(\Gamma)\implies w1\not\in p(\Gamma)$.
Suppose that $|\omega0|=n+1$ and that $\omega0=p(\omega')$ for some $\omega'\in\Gamma$
represented by the structure $\A$. Each bit in $\omega0$ corresponds to a $k$-tuple in
$p(\A)$, i.e. $\omega0\sim \bar u_0\bar u_2\ldots\bar u_n$ where $\bar u_j$ is $1$ iff
$\omega'\models\varphi(\bar u_j)$. Since $\bar u_n \sim 0$, $\omega'\nmodels\alpha_0(\bar u_n)$.
Consider the two cases whether $\omega'\models\alpha_\ell(\bar u_n)$ for some $1\leq\ell\leq r$, or not.

In the latter case, we can conclude that $\omega''\nmodels\alpha_\ell(\bar u_n)$ for
every $\omega''\in\{0,1\}^{|w|}$ and $1\leq\ell\leq r$ since $\alpha_\ell$, being a
numerical formula, obtains a value that only depends on the size of its input;
thus, $\omega1\not\in p(\Gamma)$.

In the former case, $\omega'\models\alpha_\ell(\bar u_n)$, for some unique $\ell$,
and $\omega'\nmodels\lambda_\ell(\bar u_n)$ since $\bar u_n\sim0$.
Thus, since $\lambda_\ell(\bar u_n)$ is a literal, some bit of $\omega'$ determines
the value 0 for $\bar u_n$.
On the other hand, observe that
\[\omega'\in\Gamma \iff p(\omega')=\omega0\in\Gamma' \iff \omega\in\Gamma\]
where the first equivalence follows since $p$ is a reduction, and the second by construction
of $\Gamma'$. Furthermore, being $p$ injective, implies that each bit in $\omega'$ determines
one bit in $\omega$. Therefore, there is a bit in $\omega'$ that determines two bits in $\omega0$:
one bit in $\omega$ and the rightmost 0. If $\omega1$ were in $p(\Gamma)$, then the same bit
in the preimage of $\omega1$ would determine the same bit in $\omega$ and the rightmost 1,
this time in an inconsistent manner. Therefore, $\omega1\not\in p(\Gamma)$.
\end{proof}

\section{Conclusions}

We have extended the canonical form proposed by Medina and Immerman to all complexity
classes characterized by fragments $\L$ closed under disjunctions, and under conjunctions
with FO. Although, Medina and Immerman's method could be generalized to other nice
classes beyond NP, it requires the formulation of ``generalized'' sentences. Our method,
on the other hand, circumvent this problem by considering the dual operator.
Additionally, it is not clear how Medina and Immerman's method could be used to find canonical
forms with respect to problems that are not ``graph'' problems, or on classes that 
do not have complete problems based on explicit graphs, e.g. PSPACE.

As for the near future, we are currently working on syntactic operators that
preserve completeness via fops for general complexity classes. This subject 
is also addressed by Medina  \cite{medina:thesis} where syntactic operators
$I:\L[\tau]\rightarrow\L[\sigma]$, that map formulae into formulae, are defined
such that if $\Psi$ characterizes a NP-complete problem, then so is $I(\Psi)$.
We think that as inverse images play a fundamental role in (mathematical) analysis,
inverse images of syntactic transformations are worth to explore. 
In our case, we look for operators $I$ such that if $I(\Psi)$ defines a
complete problem, then $\Psi$ also defines a complete problem; Nijjar also 
mention that such transformations are worth exploring \cite{nijjar:master}.
We believe that such operators could be use to establish completeness
of problems in an easier way.

\bibliographystyle{plain}
\bibliography{paper}






\end{document}